# Bispyrene functionalization drives self-assembly of graphite nanoplates into highly efficient heat spreader foils

G. Ferraro[1], M. Bernal[1], F. Carniato[2], C. Novara[3], M. Tortello[3], S. Ronchetti[3], F. Giorgis[3], A. Fina[1,*]

Dipartimento di Scienza Applicata e Tecnologia, Politecnico di Torino, Alessandria Campus, Viale Teresa Michel 5, 15121 Alessandria, Italy

Dipartimento di Scienze e Innovazione Tecnologica, Università degli Studi del Piemonte Orientale "Amedeo Avogadro", Viale Teresa Michel, 11 - 15121 Alessandria

Dipartimento di Scienza Applicata e Tecnologia, Politecnico di Torino, C.so Duca degli Abruzzi 24, 10129, Torino, Italy

*corresponding author: alberto.fina@polito.it

## Abstract

Thermally conductive nanopapers fabricated from graphene and related materials are currently showing a great potential in thermal management applications. However, thermal contacts between conductive plates represent the bottleneck for thermal conductivity of nanopapers prepared in the absence of a high temperature step for graphitization. In this work, the problem of ineffective thermal contacts is addressed by the use of bifunctional polyaromatic molecules designed to drive self-assembly of graphite nanoplates (GnP) and establish thermal bridges between them. To preserve the high conductivity associated to defect-free $sp^2$ structure, non-covalent functionalization with bispyrene compounds, synthesised on purpose with variable tethering chain length, was exploited. Pyrene terminal groups granted for a strong π-π interaction with graphene surface, as demonstrated by UV-Vis, fluorescence and Raman spectroscopies. Bispyrene molecular junctions between GnP were found to control GnP organization and orientation within the nanopaper, delivering significant enhancement in both in-plane and cross-plane thermal diffusivity. Finally, nanopapers were validated as heat spreader devices for electronic components, evidencing comparable or better thermal dissipation performance than conventional Cu foil, while delivering over 90% weight reduction.

## Results and Discussion

Aiming at the non-covalent crosslinking of GnP, bifunctional molecules able to provide a sufficiently strong surface interaction with graphene layers were designed and synthesised. Among the different chemical species able to strongly interact with the $sp^2$ carbon surface via π-π stacking, pyrene was selected based on its well-known adsorption on the graphene surface and its good solubility in organic solvents. A series of bispyrene (BP) molecules (referred to as 2a – 2e) were synthesized (Figure 1), varying the length of the alkyl chain linked to the pyrene units.

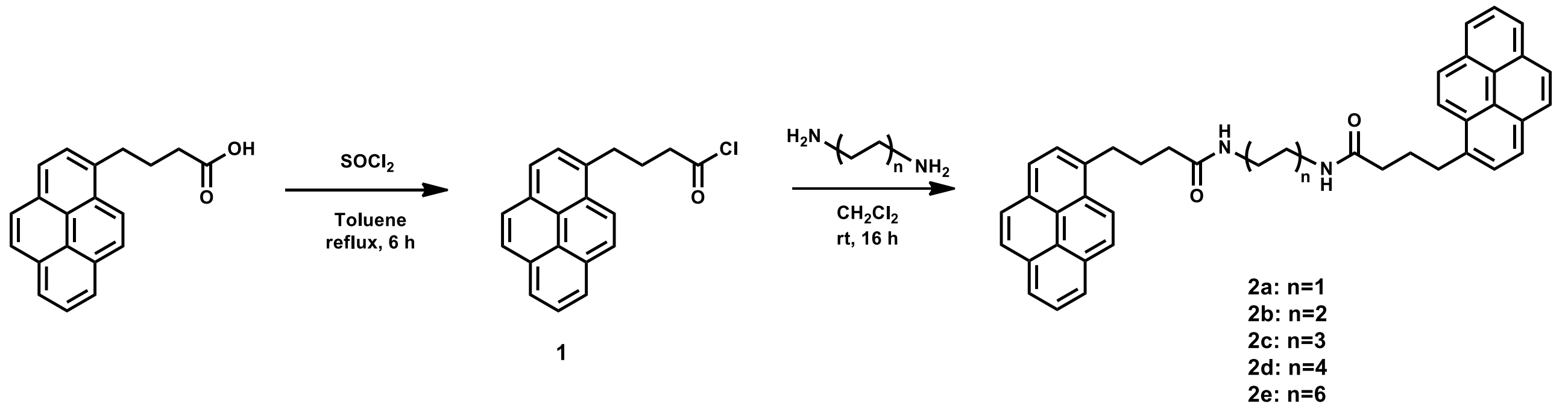

*Figure 1: Synthesis of bispyrene molecules 2a – 2e.*

The interaction of BP molecules with GnP in suspension was investigated as function of the alkyl chain length between the pyrene units, by the analysis of supernatants obtained after mild sonication of GnP in solutions of BPs. Initially, the effect of BP concentration was explored by UV-Vis and fluorescence spectroscopy to verify absorption and emission of these compounds on GnP surface, while avoiding large excess of free molecules in the systems. A BP concentration of $10^{-6}$M was selected to prevent extensive self-aggregation of BP molecules into aggregates on the surface of the nanoflakes. By UV-Vis absorbance spectra (Figure 2a), a rough estimation of the concentration of GnP suspended in supernatant can be obtained by analysing the light absorption at 670 nm, wavelength at which BPs do not absorb. The concentration of suspended GnP in the presence of BP molecules is similar or lower to that of pristine GnP, suggesting that the presence of BPs does not stabilize GnP in suspension but rather promotes aggregation of GnP, acting as π-π gelators, as previously reported for similar systems[1]. Spectrum for GnP 2a supernatant, in which reduction in GnP concentration is remarkable, clearly shows characteristic absorption bands of pyrene groups in the 300-450 nm range[2], completely absent in the spectra of supernatants containing longer BP. These facts suggest 2a to be the most efficient BP in promoting the aggregation of GnP, at the given concentration. Fluorescence emission spectra of BP in DMF solution ($10^{-6}$ M, Figure 2b) show distinctive features for both monomer (M, at 376, 396 and 418 nm) and excimer (E, at 485 nm) emissions, with variable $I_E/I_M$ intensity ratio (Figure 2b), suggesting highest intra-molecular aggregation for BP with intermediate chain lengths (2b and 2c). Fluorescence emission spectra for GnP BP supernatants (Figure 2c) show a strong reduction of $I_E/I_M$ for all BPs, which is consistent with a strong BP/GnP interaction and consequent reduction in concentration of free BP.

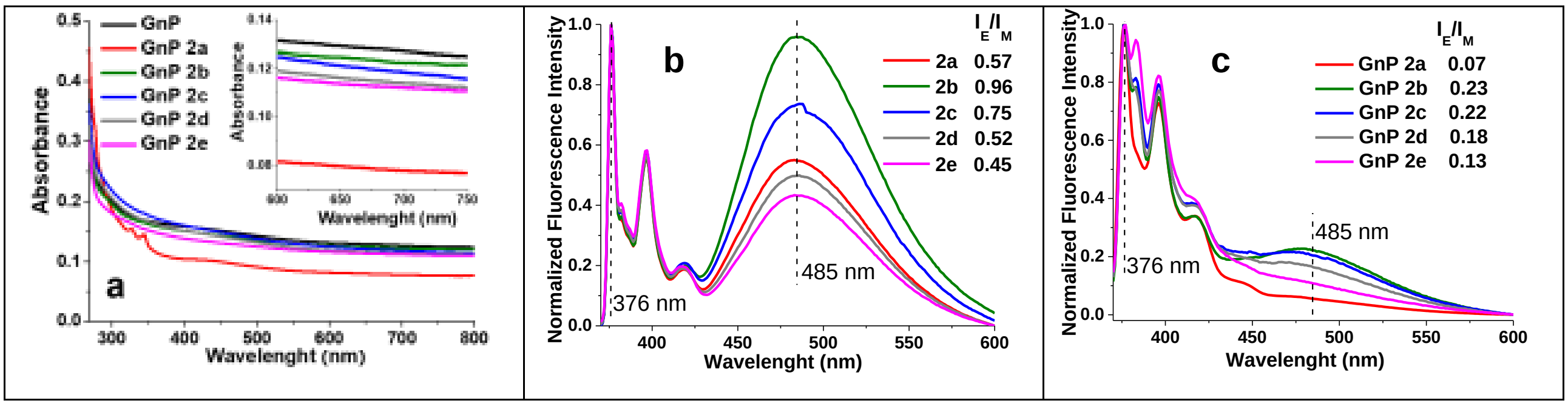

*Figure 2: UV-Vis spectra of supernatants (a) inset showing a magnification for the range 600 to 750 nm; fluorescence spectra for BP solutions in DMF, $10^{-6}$ M (b) and fluorescence spectra for supernatant of GnP/BP dispersions (c).*

The collected GnP/BP precipitates were thoroughly washed and dried, to obtain a series of supramolecular functionalized GnP. Raman spectroscopy was carried out to investigate vibrational changes, which may be correlated to the absorption of BP. In particular, the D band (approx. 1340 $cm^{-1}$), activated by graphene structural defects by second-order Raman scattering processes, the G band (around 1580 $cm^{-1}$), due to first-order Raman scattering (degenerate $E_{2g}$ mode), and in particular the ratio between their intensities ($I_D/I_G$) can be taken as a parameter for quantifying structural disorder[3-4]. The low $I_D/I_G$ ratio (0.06) of pristine GnP evidences for a low defectiveness of these nanoflakes (Figure 3). Limited increase in $I_D/I_G$ were found for BP-functionalized GnP (Figure 3), which is consistent with the π-π interaction of pyrene onto the GnP basal plane, weakly affecting the vibrational properties. Nonetheless, a slight increase in the $I_D/I_G$ ratio and its experimental deviation is observed with increasing the length of the BP, suggesting differences in the BP/GnP interaction as function of the aliphatic chain length between pyrene moieties.

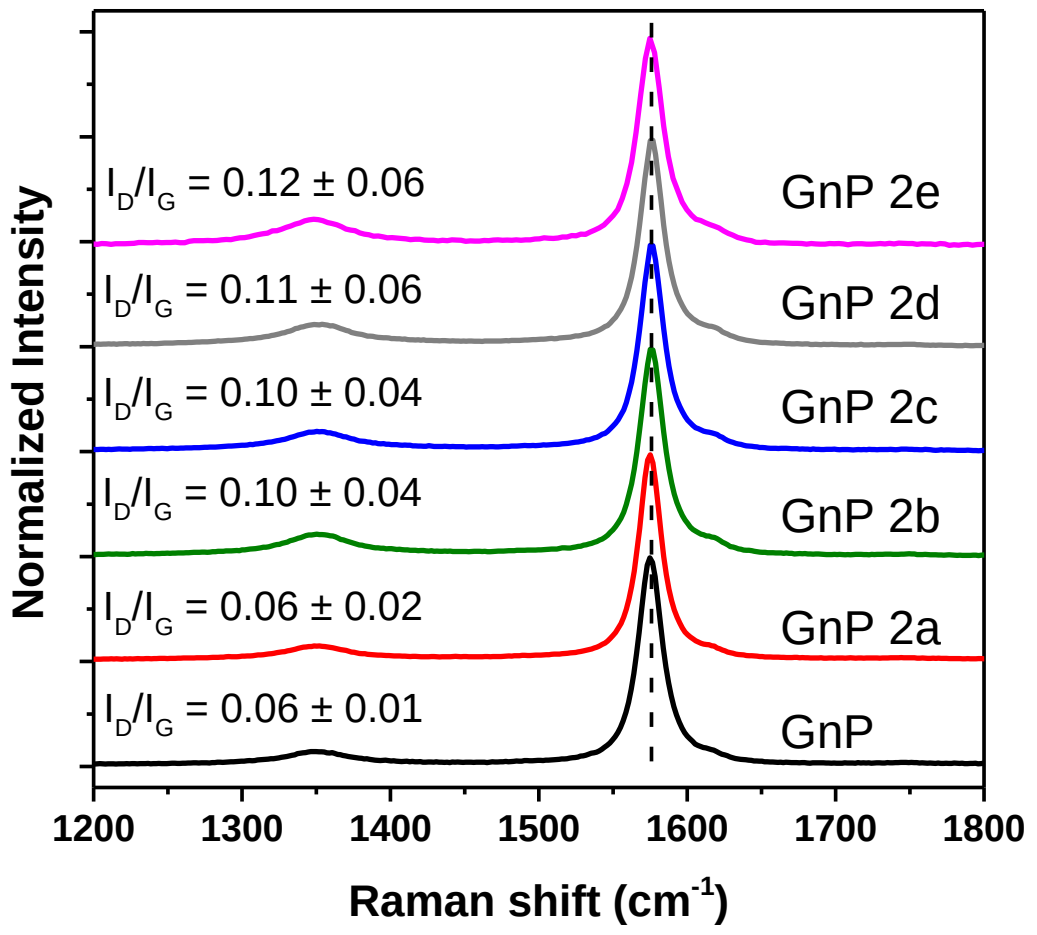

*Figure 3: Raman spectra of GnP and BP- functionalized GnP. $I_D/I_G$ band ratio are reported for each of the GnP.*

Based on the above evidence for strong adsorption of BP and promoted interactions between GnP flakes, nanopapers were fabricated by gravimetric filtration of GnP-BP suspensions, with the aim of exploiting bispyrenes to drive self-assembly between GnP flakes. The amount of BP retained within the nanopapers after filtration and washing was calculated from the concentration of BP infiltrated solutions, compared to the initial concentration in the GnP/BP suspension. Retention rate of BP into the nanopapers (Table 1) was found to be ≥95%, except in the case of GnP 2a (80%, reflecting the higher concentration of 2a in the supernatant). The mass fraction of BP in the nanopapers is ranging between 0.09 % for 2a and 0.15 % for 2e, which increase may be partially explained by the higher molar mass of longer chain BP.

*Table 1: Extent of functionalization and physical properties for GnP and BP-GnP nanopapers.*

| Nanopaper | BP Retention rate [%] | BP Mass fraction [%] | Density [g $cm^{-3}$] | In-plane Thermal diffusivity [$mm^2s^{-1}$] | Cross-plane Thermal diffusivity [$mm^2s^{-1}$] |
|---|---|---|---|---|---|
| GnP | - | - | 1.22 ± 0.05 | 175 ± 11 | 0.4 ± 0.1 |
| GnP 2a | 80 ± 1 | 0.09 ± 0.01 | 0.62 ± 0.02 | 204 ± 10 | 2.2 ± 0.2 |
| GnP 2b | 95 ± 1 | 0.12 ± 0.01 | 0.67 ± 0.02 | 192 ± 10 | 2.3 ± 0.2 |
| GnP 2c | 95 ± 1 | 0.13 ± 0.01 | 0.75 ± 0.02 | 168 ±13 | 1.5 ± 0.3 |
| GnP 2d | 99 ± 1 | 0.13 ± 0.01 | 0.77 ± 0.02 | 167 ± 15 | 1.3 ± 0.7 |
| GnP 2e | 99 ± 1 | 0.15 ± 0.01 | 0.94 ± 0.02 | 172 ± 10 | 0.6 ± 0.2 |

Such low mass fractions depend on the limited surface area of GnP (38 $m^2$/g, for the as received powder), as well as the low concentration of BP. These values correspond to a limited coverage of the GnP with BP moieties, which was designed to provide sufficient GnP functionalization, while avoiding self-aggregation of BP into the nanopapers, which is indeed possible at higher concentration. Densities of nanopapers were found to be significantly different form the reference GnP nanopaper (1.22 g/$cm^3$). In fact, lower densities were obtained for all of the GnP-bispyrene nanopapers, ranging between 0.62 and 0.94 $gcm^{-3}$. Interestingly, density values continuously increase with increasing the length of the BP alkyl chain in GnP-bispyrene nanopaper (Table 1). This trend is not straightforwardly explained in terms of BP mass fraction. Instead, it appears that the density of the GnP-bispyrene nanopaper depends on the interactions between BPs and GnP, driving the self-assembling of GnP during filtration.

To further investigate the assembly of nanopapers, X-ray diffraction (XRD) was exploited to evaluate the orientation of flakes into the films, monitoring the intensity of the graphite (002) diffraction peak at 26.6° 2θ, as a function of X-ray beam incident angle. Intensities were normalized to obtain a distribution of probability for nanoflakes orientation, as a function of tilt angle, from 0° (002 basal planes parallel to the nanopaper plane) to 90° (perpendicular to the plane). While the filtration of GnP suspensions is obviously expected to produce a clear in-plane orientation of GnP flakes, quantifying the distribution of the basal planes orientations may indeed provide insight in the assembly of GnP. The orientation for 002 planes in GnP (Figure 4a) is clearly maximum in the direction parallel to the nanopaper (0° tilt angle) for all nanopapers. However, the maximum probability values and the intensity decay profile with the tilt angle are different for GnP and the GnP-BP nanopapers. The cumulative distributions (Figure 4b) provide a quantification of preferential orientation. For pristine GnP, approx. 92% of the cumulative distribution of flakes has orientation between 0 and 10° tilt angle (i.e. an arrangement almost parallel to the surface of the nanopaper), whereas for GnP-bispyrene nanopaper in the same tilt angle range, cumulative distribution vary from 80 to 88 %. Despite no clear trend can be drawn as a function of the tethering chain length in BP, these values evidence for a slightly lower in-plane orientation of the GnP-bispyrene nanopaper, compared to pristine GnP, confirming BPs to have a role in the self-assembly of GnP during filtration. This is likely explained by the formation of BP-bridged aggregation of GnP flakes in the suspensions into clusters, which in turn constraint the organization of flakes during filtration, decreasing the packing factor and eventually reducing preferential orientation and density.

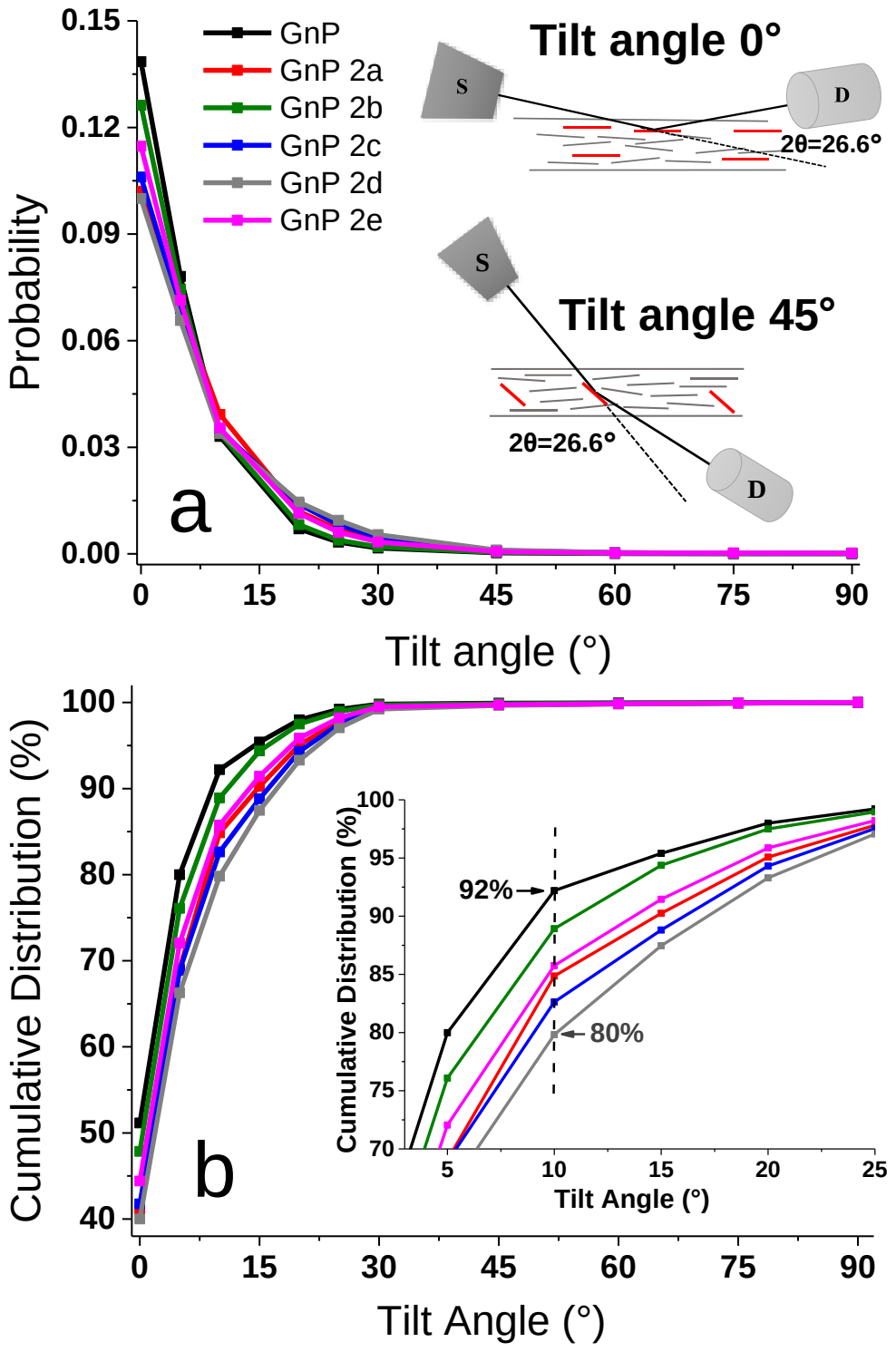

*Figure 4: a) Probability distribution for the orientation of GnP flakes, based on the 002 diffreaction signal for basal planes in graphite. Inset shows a schematic representation of configuration used, as a function of indicident angle, while maintaining constant 2θ=26.6°. In red are highlighted GnP flakes contributing to the diffracted beam, oriented at 0° and 45° tilt angles, respectively. S is for X-ray source, D is for X-ray detector. b) Cumulative distribution for GnP orientation vs. tilt angle. Inset shows a magnification in the tilt angle range 3 to 25°.*

Insight on the heat spreading efficiency of nanopapers was obtained by measuring thermal diffusivity, i.e. the rate of heat transfer within the porous material. Both in-plane and cross-plane diffusivities were measured for pristine GnP and GnP-bispyrene nanopapers (Table 1 and Figure 5a), evidencing for significant differences induced by BP functionalization. As expected from the orientation of GnP, thermal diffusivity in nanopapers is strongly anisotropic, with at least two orders of magnitude higher diffusivity in plane, compared to cross-plane. Cross-plane thermal diffusivity for GnP/BP nanopapers are typicslly higher than that of pristine GnP, which may be partially related to the above described lower nanoflakes orientation. Indeed, the presence of even very limited portions of flakes lying on directions tilted with respect to the nanopaper plane is expected to contribute to the heat transfer across the nanopaper, owing to the well-known anisotropy of graphitic materials. However, the changes in cross-plane diffusivity do not match misalignment degree, as evaluated in XRD, suggesting a different explanation about the role of BP moieties. Short BP (2a and 2b) resulted to be the most effective functionalization for the in-plane diffusivity enhancement, suggesting these molecules to reduce the thermal resistance between overlapped nanoflakes. This may be explained by promoted aggregation between GnP, maximizing their contact area, as well as by the enhancement of phonon transfer efficiency through BP molecules bridging adjacent GnP, thus acting as a non-covalent molecular junctions for phonon transfer. BP 2a and 2b also significantly enhanced in-plane thermal diffusivity, whereas values for GnP-bispyrene nanopapers containing longer BPs are equivalent to pristine GnP. The in-plane diffusivity enhancement is obtained in spite of the lower in-plane orientation for GnP-bispyrene nanopaper, thus further supporting the role of BP molecular junctions in the reduction of thermal resistance at GnP-GnP contacts. It is also worth mentioning that the described trend in thermal diffusivity is not corresponding to significant changes in volumetric electrical conductivity for the nanopapers, measured in the range of $2 \cdot 10^5$ S/m, highlighting the effect of BP molecules is specifically on thermal transport properties.

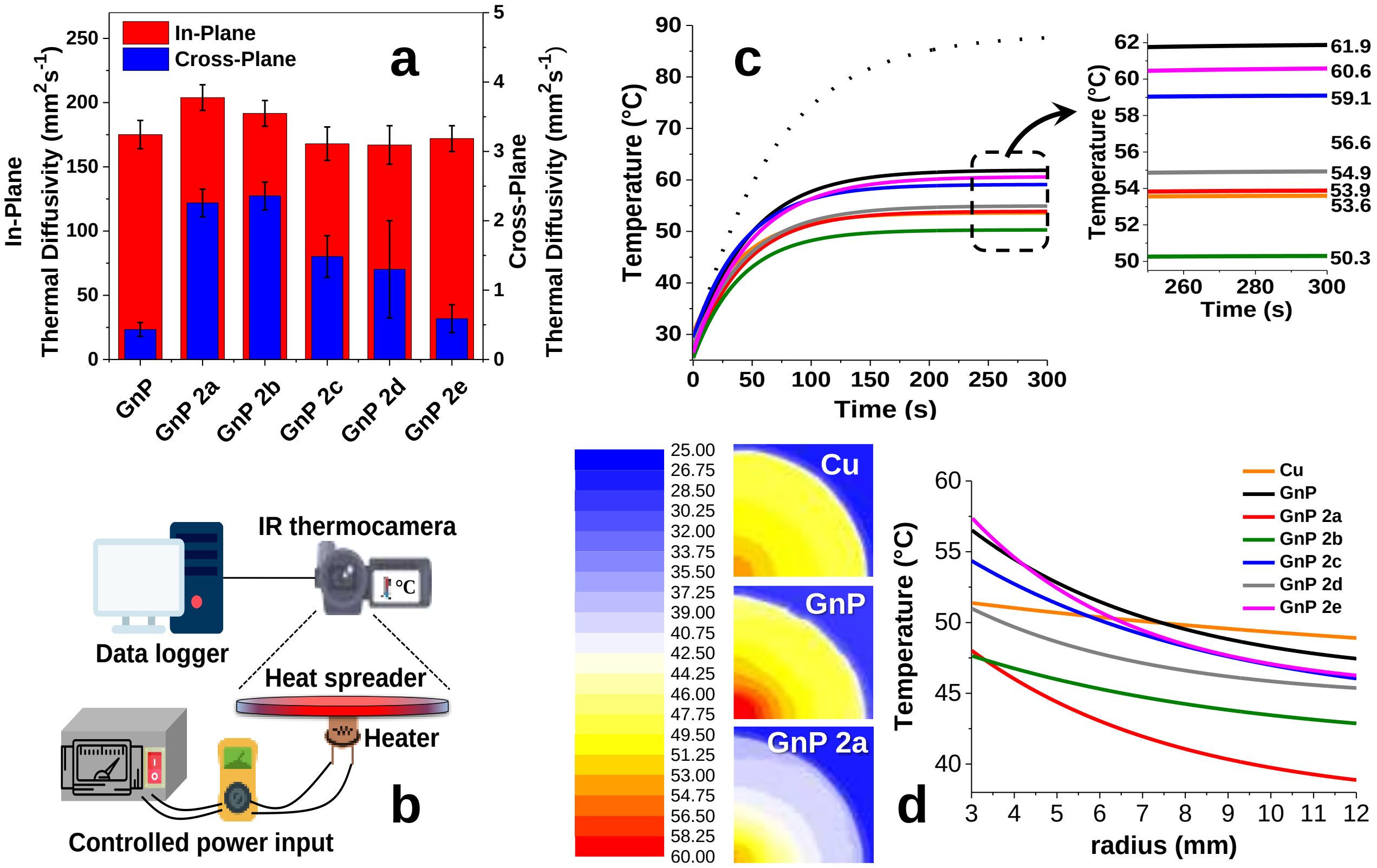

*Figure 5: Thermal properties of GnP nanopapers. a) In-plane and cross-plane thermal diffusivities for GnP and GnP-bispyrene nanopaper; b) Setup for the measurement of thermal properties as heat spreader; c) Hotspot temperature profile vs time for GnP nanopapers vs. copper foil d) Thermal imaging at 300 seconds of selected heat spreaders (I quadrant only for better visibility) and fitted temperature gradient vs radius comparison for GnP nanopapers vs. copper foil.*

From an application perspective, such increased in-plane thermal diffusivity may be exploited in heat spreader applications, used for the dissipation of heat from a hotspot, over a larger surface. This typically occurs in several electronic components used in modern devices, especially in high frequency processors, where the excess heat must be efficiently dissipated to prevent potential failures. Copper is traditionally used for heat spreaders, in the forms of foils or finned structures. GnP nanopapers are in principle a valid alternative to metals, owing to their flexibility and low density, thus opening for applications in flexible and lightweight devices, such as flexible electronics as well as wearable and implantable devices. With the aim of comparing performances of GnP-bispyrene nanopaper with respect to the conventional metal foil, a simple experimental setup (Figure 5b) for the measurement of heat spread from a hotspot was built. Using a powered transistor, simulating an hotspot and an IR thermocamera to collect thermal maps in time, the temperature evolution over the heat spreader, made with copper film or with the GnP nanopapers, was continuously monitored during heating (power on, Figure 5c) and cooling (power off,supportin) of the hotspot. In the absence of a heat spreader foil, the temperature of the hotspot raises to about 85°C after 300 s power on. As expected, the presence of copper foil strongly reduced the hotspot temperature to 53.6°C @ 300s (Figure 5c, inset), as the heat is distributed on the surface of the foil and eventually dissipated to the surrounding air, by natural convection. Copper foil was used as a benchmark, based on its well-known thermal conductivity in the range of 400 $Wm^{-1}K^{-1}$. With nanopaper heat spreaders, temperature profiles vs. time were found strongly dependent on the presence and type of functionalization (Figure 5c). Indeed, while all GnP-bispyrene nanopaper have a maximum hotspot temperature lower than with the pristine GnP, Nanopaper GnP 2a revealed to be as effective as Cu foil and GnP 2b demonstrated even better performance, allowing to maintain a hot-spot temperature (50.3 @300s) constantly lower than in the presence of copper foil. This outstanding result is even more impressive when considering the densities of the heat spreaders, as the porous GnP 2b nanopaper is only 0.67 $gcm^{-3}$, whereas copper is about 8.9 $gcm^{-3}$, thus delivering a straightforward weight reduction of the heat spreader by one order of magnitude.

To further investigate the heat dissipation over the spreader foils, the thermal maps acquired in time, upon heating and cooling the transistor, were analysed to calculate the thermal gradient on the foil *vs.* the radial coordinate. Selected temperature maps reported in Figure 5d clearly evidence for significant differences in the temperature gradient over the surface for copper, GnP and BP-GnP. In particular, a lower temperature gradient is apparent on Cu foil, compared to GnP nanopapers, evidencing for an efficient distribution of heat over the surface, according to its high thermal conductivity value. Comparison between temperature vs. radius profiles for Cu and GnP nanopapers (Figure 5d, Figures S31-S33) confirms a systematically higher temperature decay rate on the GnP foils, suggesting a lower capability to distribute the heat flow over the nanopapers, likely related to their lower density. These results may appear contradictory to the thermal diffusivity results, but it has to be recalled that the heat flux is dependent on thermal conductivity, which in turn depends linearly on the density of the dissipating materials. Given the porosity of the nanopaper is relatively high, the heat spreading capability is indeed expected to be lower than for a high-density metal. Therefore, the demonstrated performance of GnP nanopapers to keep the hotspot at low temperature appears to be related to phenomena beyond the bare heat conduction. In particular, the overall performance of the heat spreader may also strongly depend on thermal transfer at the solid/air interface, which is directly associated to the porosity and surface morphology of GnP nanopapers, eventually leading to a heat dissipation performance comparable to or even higher than copper foil.

## Conclusions

Graphite NanoPlates (GnP) were functionalized with a new family of BisPyrene (BP) molecules, designed to anchor pyrene moieties on GnP and connect them via an aliphatic chain with variable length. Synthesised BP demonstrated strong adsorption onto GnP and tendency to promote aggregation of nanoflakes in suspension, thus suggesting potential in the self-assembly of GnP. Functional GnP foils were manufactured via gravimetric filtration method and referred to as nanopapers, owing to similarities with the well-known paper-making process.

GnP organization within nanopapers was evaluated by electron microscopy and X-ray diffraction, to quantify porosity and orientation of the flakes within the nanopapers, demonstrating a strong role of BP in the self-assembly of GnP, leading to lower densities and reduced preferential in-plane orientation. Thermal diffusivity of GnP nanopapers were found to depend on the length of BP, with shortest BPs delivering best thermal diffusivity in both cross-plane and in-plane diffusivities. Indeed, while GnP nanopaper showed 175 $mm^2s^{-1}$ in plane and 0.4 $mm^2s^{-1}$ cross plane thermal diffusivity, GnP 2a nanopapers exhibited thermal diffusivity values of 204 $mm^2s^{-1}$ and 2.2 $mm^2s^{-1}$ for in-plane and cross-plane diffusivity. These evidence support for promoted aggregation between GnP, maximizing their contact area and enhancing phonon transfer efficiency within the network of conductive flakes, demonstrating bispyrene molecules as effective drivers for a thermally efficient self-assembly of GnP.

To demonstrate the performance of GnP nanopapers in conditions representative of actual heat spread applications, a simple experimental setup was used to simulate cooling of an overheated electronic device (as a representative hotspot). In these conditions, GnP nanopapers were benchmarked to a copper foil with the same geometry. Hotspot temperature measures demonstrated all BP-functionalized nanopapers perform better than pristine GnP nanopapers. Most importantly, one functionalized nanopaper demonstrated better performance in cooling the hotspot, compared to the copper benchmark. Beside thermal performance, a dramatic 90% weight reduction was obtained for the nanopapers, compared to copper foil, which may open for potential application of thermally conductive nanopapers in a number of lightweight and flexible devices.

## Materials and Methods

Graphite nanoplates (GnP) with a mean size of some tens of micrometres and a thickness of few nanometres were prepared and provided by Avanzare Innovación Tecnólogica S.L (Navarrete, La Rioja, Spain) via rapid thermal expansion of overoxidized-intercalated graphite, according to a previously reported procedure[5]. Detailed characterization of this GnP was previously reported [5-6].

**Synthesis of bispyrene molecules.** 1-pyrenebutyric acid (300 mg, 1.4 mmol) (Sigma Aldrich, 97%) was dissolved in 100 mL dry toluene (Sigma Aldrich, anhydrous, 99.8%) and cooled at 5 °C. Thionyl chloride (Sigma Aldrich, ≥99.0 %) (10 mL) was added drop wise and the mixture was heated to reflux for 6 h under an argon atmosphere. The solvent was evaporated in vacuum, obtaining an orange solid, which was subsequently dissolved in dichloromethane (20 mL) (Sigma Aldrich, anhydrous, ≥99.8%). A solution of the diaminoalkane (0.5 mmol, 1,2-Diaminoethane, Sigma Aldrich, ≥99.5%; 1,4-Diaminobutane, Sigma Aldrich, ≥99%; 1,6-Diaminohexane, Sigma Aldrich, ≥99%; 1,8-Diaminooctane, Sigma Aldrich, 98%; 1,12-Diaminododecane, Sigma Aldrich, 98%) in dichloromethane (2 mL) was added drop wise to the solution containing 1-pyrenebutyric acid and the reaction mixture was stirred at room temperature for 18 h. The precipitate was filtered using nylon filter membranes (pore size 0.45 µm, Whatman), washed with dichloromethane and the solvent was evaporated. In order to remove the non-reacted diaminoalkane, the solid was stirred for 1 h in a $10^{-4}$ M solution of hydrochloric acid (Sigma Aldrich, 37%) and filtered to obtain the final bispyrene (BP) product. Nuclear Magnetic Resonance Analysis (NMR) were acquired on Bruker AVANCE III (Italy) at 500 MHz (≈ 3 mg of powder were solubilized into 1 ml of hexadeuterodimethyl sulfoxide, Sigma Aldrich, 99.96 atom % D). Liquid chromatography coupled with mass spectrometry (LC-MS) were used to acquire the spectra (samples < 1 mg solubilized into 5 ml of methanol for HPLC, ≥99%) with an electrospray ionization $H^+$ mode (model LTQ XL, Thermo Fisher Scientific, USA).

**Functionalization of GnP with BP.** GnP (10 mg) were added to a solution (20 mL of the selected BP in N,N-dimethylformamide ($10^{-6}$ M) (Sigma Aldrich, anhydrous, 99.8%) at a concentration of 0.5 mg $mL^{-1}$. The solutions were sonicated in pulsed mode (5 s on/5 s off) for 30 min at a power output of 150 W by using an ultra-sonication probe (Sonics Vibracell VCX-750, Sonics & Materials Inc, USA) with a 5 mm diameter Ti-alloy tip. After that, the suspension was left to decant for 120 min at 5°C and part of the supernatant was carefully collected for further characterization, while the other part was filtrated through a polytetrafluoroethylene (PTFE) supported membrane (0.2 µm as pore size, Whatman), washed with deionized water (50 mL), ethanol (50 mL, Carlo Erba, 96%) and diethyl ether (50 mL, Sigma Aldrich, anhydrous, ≥99.7%) and dried at 60 °C for 24 hours. Raman spectra were acquired on a Renishaw inVia Reflex (Renishaw PLC, UK) Raman microscope using an excitation laser wavelength of 514.5 nm. The measurements were performed using a 20x objective in backscattering configuration, setting the laser power to 2.5 mW and the integration time to 50 s. The reported spectra are the average of five measurements acquired in different areas of the sample. Peaks have been fitted with a Lorentzian function using the origin Pro 2016 software. The supernatant was centrifuged at 4000 rpm for 30 min and left to decant overnight. Finally, the part above the sediment (supernatant) was sampled and analysed by UV-Visible spectroscopy, (UV-2600, Shimadzu, Japan) with a single scan with a 0.5 nm sampling interval and 0.05 s accumulation time, 1 cm quartz cuvette, and the photoluminescence spectra were recorded on a Horiba Jobin-Yvon Model IBH FL-322 Fluorolog 3 spectrometer equipped with a 450 W Xenon arc lamp with excitation at 345.00 nm.

**Preparation and characterization of GnP-BP nanopapers.** GnP and BP-functionalized GnP were suspended on N,N-dimethylformamide (Sigma Aldrich, anhydrous, 99.8%) solutions (0.5 mg $mL^{-1}$) and sonicated in pulsed mode (5 s on and 5 s off) for 30 min with power set at 150 W) by using an ultra-sonication probe (Sonics Vibracell VCX-750, Sonics & Materials Inc, USA) with a 13 mm diameter Ti-alloy tip. The suspensions were gravimetrically filtrated using a polyamide membrane (0.45 µm as pore size, Whatman) and then fixed in a petri glass with scotch tape to be dried at 65 °C under vacuum for 2 h to completely remove the solvent. Filtrated solutions were collected and analysed by UV-Vis (UV-2600, Shimadzu, 1 cm quartz cuvette with single scan 0.5 nm sampling interval and 0.05 s accumulation time) to measure the BP absorbance and calculate the concentration from relative calibration line using $y=m \cdot x$ equation (where y is the absorbance, m is the slope of the curve and x is the concentration of the solution). Mass fraction of BP adsorbed onto GnP was calculated from the difference between initial concentration and concentration in the filtered solution.

Nanopapers were peeled off from the membranes and were mechanically pressed in a laboratory hydraulic press (Atlas 15T, Specac, England) under a uniaxial compressive load of 5 kN for 30 min at 25 °C. Nanopapers were die cut into 0.53 mm diameter disks to calculate the densities, as the ratio between mass measured by a microbalance (resolution 1 µg, TA Discovery TGA used at roome temperature) and the volume calculated from the known diameter and the thickness, measured by field-emission scanning electron microscopy (FESEM, Zeiss Merlin 4248, beam voltage: 5kV, Germany) on the cross-section of the GnP nanopapers.

Orientation of GnP in the nanopapers were evaluated using an X-ray Panalytical X'Pert PRO (Cu Kα radiation) diffractometer, with a PIXcel detector, a solid-state detector with rapid readout time and high dynamic range. The specimens (disks 23 mm diameter, 20 to 45 µm thick) were supported onto PET film (5 µm) and measured with a 2theta scan axis for each value of the incident angle (α) ranging from 13° to 103° with respect to the horizontal plane of the sample. Intensities of the signal of 002 graphite planes (2θ=26.6°) were collected against the incident angle. Intensities are reported as a function of the tilt angle $\alpha$-$\theta_{002}$, which corresponds to the tilt angle of the flakes in the nanopaper, varying from ~0° (basal planes parallel to the nanopaper plane) to ~90° (basal planes perpendicular to the nanopaper plane). The intensity *vs.* 2θ experimental curves (for each incident angle) were fitted with a Lorentzian function using the OriginPro 2020 software in order to obtain the peak intensity for 002 diffraction signal. Then, the peak intensities vs tilt angle were plotted for each of the nanopapers and fitted by an exponential decay curve using OriginPro 2020 software. Finally peak intensities vs tilt angle plots were normalized on their integral value, to obtain the probability distribution for the flakes oriented from parallel to perpendicular to the nanopaper plane.

The in-plane thermal diffusivity ($\alpha_{\parallel}$) and cross-plane diffusivity ($\alpha_{\perp}$) were measured at 25 °C using the xenon light flash analysis (LFA) (Netzsch LFA 467 Hyperflash, Germany). The samples were die cut in disks of 23 mm with typical thicknesses in the range 20 – 50 μm and the measurement of the $\alpha_{\parallel}$ was carried out in an in-plane sample holder while the $\alpha_{\perp}$ was measured in the standard cross-plane configuration. At least three different specimens were tested for each of the nanopaper formulations. Five measurements were collected on each sample to calculate average values and experimental deviations.

**Heat spreader device setup.** Heat spread performance was evaluated using the same specimen geometry used for LFA, assembled onto a transistor (2N2222 A, low power bipolar transistor, NPN silicon planar switching transistor, TO-18 metal package, STMicroelectronics), powered by an electrical generator (GWinstec GPS-3303, Taiwan), set at 0.080 A current at 4.3 V to provide 0.345 W. The thermal contact of the nanopaper disk heat spreader onto the flat surface (12.56 $mm^2$) of the transistor is mechanically guaranteed by a 3x3x3$mm^3$ NdFeB (N28) magnet, centred onto the transistor top surface. The heat spreader disks, as well as the magnetic cube were black paint sprayed to control their emissivity. Temperature of the system was monitored by an IR thermal imaging camera (Optris-Cam PI-400, Germany), with an optical resolution of 382 x 288 pixel and 1 Hz sampling rate. Tests were carried out under a closed box measuring 29x23x19 $cm^3$ to ensure the reproducibility of natural convection upon heating (power on) for 300 second and cooling (power off), for other 300 sec. Thermal maps on the magnet surface were elaborated using OriginPro 2020 software. Hotspot temperature was calculated as the average temperature over the 3x3$mm^2$ area in top of the magnet, which is directly related to the temperature of the transistor. Temperature profiles *vs.* radial coordinate along three different directions (considering a Cartesian x, y system with the origin centred in the centre of the sample, the horizontal, vertical and oblique at 45° directions of the first quadrant) were extracted and averaged to obtain a representative temperature *vs.* radius decay curve.

**Acknowledgment**

This work has received funding from the European Research Council (ERC) under the European Union’s Horizon 2020 research and innovation programme, Grant Agreement 639495 — INTHERM — ERC-2014-STG.

The authors gratefully acknowledge Julio Gomez at Avanzare Innovación Tecnólogica S.L (E) for kindly providing GnP and Mauro Raimondo at Politecnico di Torino-DISAT for Electron microscopy. Special thanks to Fausto Franchini at Politecnico di Torino-DISAT for his precious assistance in the heat spreader demonstrator setup and measurements.

## Competing Interest

The authors declare no competing interest.

## Author contributions

A.F conceived this research work and the experiments within, interpreted the experimental results and led the research project. G.F. carried out the synthesis of BP, GnP functionalization, nanopapers preparation and most of the characterization. M.B. contributed to the synthesis, elaboration and interpretation of results. F.C. carried out and interpreted florescence, NMR and Mass Spectroscopy tests. C.N. and F.G carried out Raman analyses and interpretation. S.R. contributed to the design of XRD test and carried out measurements. M.T. carried out electrical resistivity measurements and interpretation. Manuscript was mainly written by A.F. and G.F.